\documentclass{aa}
\usepackage{graphicx,lscape,}
\usepackage{txfonts}
\begin{document}
\title{Trigonometric parallaxes of high velocity halo white dwarf candidates
\thanks{Based on observations collected at the European Southern Observatory, Chile (067.D-0107, 069.D-0054, 070.D-0028, 071.D-0005, 072.D-0153, 073.D-0028) }
}

\author{C.~Ducourant \inst{1}
       \and
         R.~Teixeira \inst{2,1} 
              \and
        N.C.~Hambly \inst{3}
              \and
        B.~R.~Oppenheimer \inst{4}
               \and
        M.R.S.~Hawkins \inst{3}
              \and
        M.~Rapaport \inst{1}
              \and
        J.~Modolo \inst{1}
              \and
       J.F.~Lecampion \inst{1}}

  \institute{Observatoire Aquitain des Sciences de l'Univers, CNRS-UMR 5804, BP 89, 33270 Floirac, France.
     \and
        Instituto de Astronomia, Geof\'isica e Ci\^encias Atmosf\'ericas,
        Universidade de S\~ao Paulo,
        Rua do Mat\~ao, 1226 - Cidade Universit\'aria,
        05508-900 S\~ao Paulo - SP,
        Brasil. 
    \and
        Scottish Universities Physics Alliance (SUPA), Institute for Astronomy,
        School of Physics, University of Edinburgh, Royal Observatory,
        Blackford Hill, Edinburgh, EH9~3HJ, UK.
    \and
       Department of Astrophysics, American Museum of Natural History, 
       79th Street at Central Park West, New York, NY 10024-5192, USA.
}

\offprints{ducourant@obs.u-bordeaux1.fr}

\date{Received  / Accepted }
\titlerunning {Parallaxes of halo white dwarf candidates}


\abstract
 {
 The status of 38 halo white dwarf candidates identified by Oppenheimer et 
al.~(2001) has been intensively discussed by various authors. In analyses 
undertaken to date, trigonometric parallaxes are crucial missing data.
Distance measurements are mandatory to 
kinematically segregate halo object from disk objects and hence enable
a more reliable estimate of the local density of halo dark matter residing in 
such objects.}
  {
  We present trigonometric parallax measurements for 15 candidate halo 
  white dwarfs (WDs) selected from the Oppenheimer et al.~(2001) list.
  }
   {
   We observed the stars using the ESO~1.56-m Danish Telescope and 
ESO~2.2-m telescope from August 2001 to July 2004.
   }
 {
 Parallaxes with accuracies of 1--2 mas were determined 
 yielding relative errors on distances of $\sim5$\% for 6 objects, 
$\sim12$\% for 3 objects, and $\sim20$\% for two more objects. 
Four stars appear to be too distant (probably farther than 100~pc) to have
measurable parallaxes in our observations.
 }
 {
Distances, absolute magnitudes and revised space velocities were derived 
for the 15 halo WDs from the Oppenheimer et al.~(2001) list.  
Halo membership is confirmed unambiguously for 6 objects 
while 5 objects may be thick disk members and 4 objects are too distant to 
draw any conclusion based solely on kinematics. Comparing our trigonometric 
parallaxes with photometric parallaxes used in previous work reveals an 
overestimation of distance as derived from photometric techniques. 
This new data set can be used to revise the halo white dwarf space density,  
and that analysis will be presented in a subsequent publication.

} 

\maketitle


\section{Introduction}
In the last decade interest in the very cool, old white dwarf
(WD) halo population has grown. This interest is motivated by the 
possibility that these objects could account for a significant 
fraction of the baryonic dark matter of our Galaxy.
This idea is in accord with discussions 
attempting to explain the microlensing events in the Large Magellanic Cloud 
in terms of a halo WD population -- see, for example, \cite{chab96} and 
\cite{hans98}. \cite{alco99} suggested that 
massive compact halo objects (MACHOs) make up~20 to~100\% of the dark matter 
in the halo, with MACHOs having typical mass $m\sim0.5$~M$_{\odot}$; more
recently, \cite{calc05} find a similar result from pixel lensing in the line
of sight to~M31. Hence, in this scenario the search for, and direct study of, 
halo WDs can provide constraints on the fraction of dark 
matter in the Milk Way that is attributable to these objects.

Oppenheimer et al.~(2001, hereafter OHDHS) 
identified 38 high proper motion WDs; 
from their kinematics, the authors concluded that they were members 
of a halo population. Since then an intense discussion 
concerning the status of these objects has taken place in the literature.
A comprehensive review of this debate is presented in \cite{hanl03} where
the conclusion is that the OHDHS interpretation is possibly overstated, 
but that complete conclusions are not possible without further data.  
Other studies suggest that the disk and ``thick disk'' Galactic populations 
can be used to explain the great majority of the objects (\cite{reid05, kili05, spag04, 
crez04, holp04, flyn03, silv02}). 
The importance of the high velocity WDs cannot be understated in other contexts 
(e.g.~the star formation history
of the Galaxy, see also \cite{davi02,hans03,mont06}). 
Moreover, several studies emphasise the importance of
obtaining trigonometric parallaxes for candidate halo WDs
(\cite{berg02,torr02,berg03}). This is especially
important for the coolest WDs, whose spectral energy distributions show
remarkable departures from black--body distributions and which are proving
to be difficult to model accurately (\cite{kowa06, gate04, saum99, hans98}).
In the presence of such radical changes to the WD spectrum, the assumption
of a monotonic photometric parallax relation (e.g.~as used in OHDHS) could
break down and estimates of intrinsic space velocities could be in
error seriously. Furthermore, a recent paper (Bergeron et al.~2005) 
concludes that precise distances are mandatory to derive accurate kinematics 
and ages for the putative halo WDs and in order to derive their 
evolutionary status.

Aiming to clear up this question, in 2001 we started an observing
program with the ESO 1.56-m Danish and ESO 2.2-m telescopes
to measure the trigonometric parallaxes of these stars. 
Trigonometric parallax 
measurements remain the only direct unbiased distance determination. 
They are of great importance in the debate about the status of cool halo 
white dwarfs because they are required to derive precise space velocities and ages 
which are used for distinguishing between halo and disk membership. 
These trigonometric parallaxes 
lead to the re-calibration of photometric distances used until now in 
this debate and allow analysis of the 
cool halo white dwarf population with more confidence.  Unfortunately, due to 
limited observing time, only 15 stars on the 
OHDHS list have been observed to date.  
However, this sub-sample provides important insight into the problem.

\section{Observations}
\label{obs}
Astrometric observations of 15 of the OHDHS list of 38 halo white dwarf 
candidates were performed at the ESO 2.2-m telescope equipped with the WFI 
wide--field mosaic camera (with 0.238 \arcsec/pixel, a field of view 
FOV = 34\arcmin~$\times$~33\arcmin, 4~$\times$~2 mosaic of 2k~$\times$~4k CCDs), 
through the ESO 
845 I filter.  To reduce astrometric distortions and other instrumental effects, only 
data from chip 51 (with FOV~=~8\arcmin $\times$ 16\arcmin) were used in this work; 
target stars were centered in the FOV of this chip.

Four epochs of observation were acquired at maximum 
parallactic factor in Right Ascension in November 2002,  July 2003, 
November 2003 and July 2004 with a total of 11 nights of observations. 
Two parallactic periods (four observations over 1.5 years) are required, 
at a minimum, for a unique determination of the parallax and proper motion. 
Two preliminary observing runs were performed at the ESO 1.56-m Danish 
telescope in July 2001 and July 2002 but the subsequent closure of the 
telescope forced the authors to move the program to the ESO 2.2-m telescope. 
Data acquired at the Danish telescope were not included in our final analysis 
to avoid systematic effects due to the use of two different telescopes. \\

To minimize differential colour refraction effects (DCR), observations were 
performed around the transit of targets with hour angles of less than 1 hour. 
Multiple exposures were taken at each observation epoch to reduce the astrometric 
errors  and to estimate the precision of measurements. Exposure times varied from 
100 to 600 seconds depending on the magnitude of the target. Each field was observed 
from 20 to 35 times.

\section{Astrometric Reduction}
\label{ast}

\subsection{Measurements}
Frames were measured using the {\tt DAOPHOT II} package 
(Stetson 1987), fitting a PSF. The significance level of a luminosity
enhancement over the local sky brightness which was regarded as
real was set to 7$\sigma$.  The PSF routine was used to define
a stellar point spread function for each frame. Finally we obtained the $(x,y)$ 
measured positions, the internal magnitudes and associated errors of all stars 
on each frame.
There were typically 300 to 600 stars measured on each frame depending on the 
exposure time. From these, a selection on the error in magnitude (ERRMAG) as 
derived by the DAOPHOT II software was applied. Any observation with ERRMAG 
$\ge 0.15^{\rm m}$ was rejected. Objects fainter than 1.5$^{\rm m}$ brighter 
than a given image's limiting magnitude were also rejected from the analysis.

\subsection{Cross-Identification}
For each of the 15 different fields of view, we selected a ``master'' or 
fiducial image from the set of 20 to 35 images.  This master frame for each object 
had the deepest limiting magnitude and highest image quality.   For each of the 
other images for a given target object, the positions of all stars not rejected 
by the criteria above were then cross--identified to the master image's star positions. 
Objects not detected on three or more frames were excluded, yielding 
100 to 200 stars in common in each field. Frames containing less than 
$N_{\rm master}/3$ stars in common with the master frame were removed from 
the solution (where $N_{\rm master}$ is the number of stars in the master 
frame). Note that the master frame is processed in an identical fashion to the other 
frames and is not assumed to be free of errors in the parallax solution.  
In other words, the fiducial frames are not taken as an error-free ``truth'', 
but are simply used as a basis for coordinate transformations and correlation 
of star positions that comprise the astrometric grid used in the solution.

\subsection{Differential Colour Refraction}
Atmospheric refraction changes the apparent positions of stars in ground--based 
observations and depends on the zenith distance of the observations.  For precision 
astrometry this effect must be accounted for, because it can be many tens of 
milliarcsececonds at even relatively modest zenith distances.
In our case, another effect becomes important as well, because the atmospheric 
refraction of our target stars will not be identical to that of the background 
stars used for our astrometric reference grid.  Our target stars (WDs) and the 
background stars (typically main--sequence G or K stars) have different
spectral energy distributions.  Therefore, atmospheric refraction will affect 
them differently when observed through a given filter bandpass.  This is called a 
differential colour refraction (DCR) and is known to cause spurious parallactic 
motion \cite{mone92}. DCR can affect both the Right Ascension (RA) and Declination of 
the target as derived with respect to the field stars. Observations in parallax 
programs are planned to maximize the parallactic factor in RA so the 
parallax solution for the target will rely heavily on the RA 
measured. Therefore the parallax derived is mainly perturbed by the DCR 
effects in RA which are critically dependent on the zenith distance of 
a given observation.

We investigated the impact of such effects on the parallax of white dwarfs 
through simulations. Using the usual formula for atmospheric refraction, a blackbody 
approximation for white dwarf and background stellar spectra, the Besan\c con 
Galaxy model for background star characteristics (\cite{robi94})
and ESO 845 filter limits, we computed the average differential colour 
refraction effects between a white dwarf similar to those of our list with 
effective temperatures, T$_{\rm eff}$, in the range 4000~K to 11000~K (\cite{berg05}, 
Table~2) and a typical background star (T$_{\rm eff} \sim 5000$ K). 

\begin{figure}[ht]
\begin{center}
\includegraphics*[width=7cm,angle=-90]{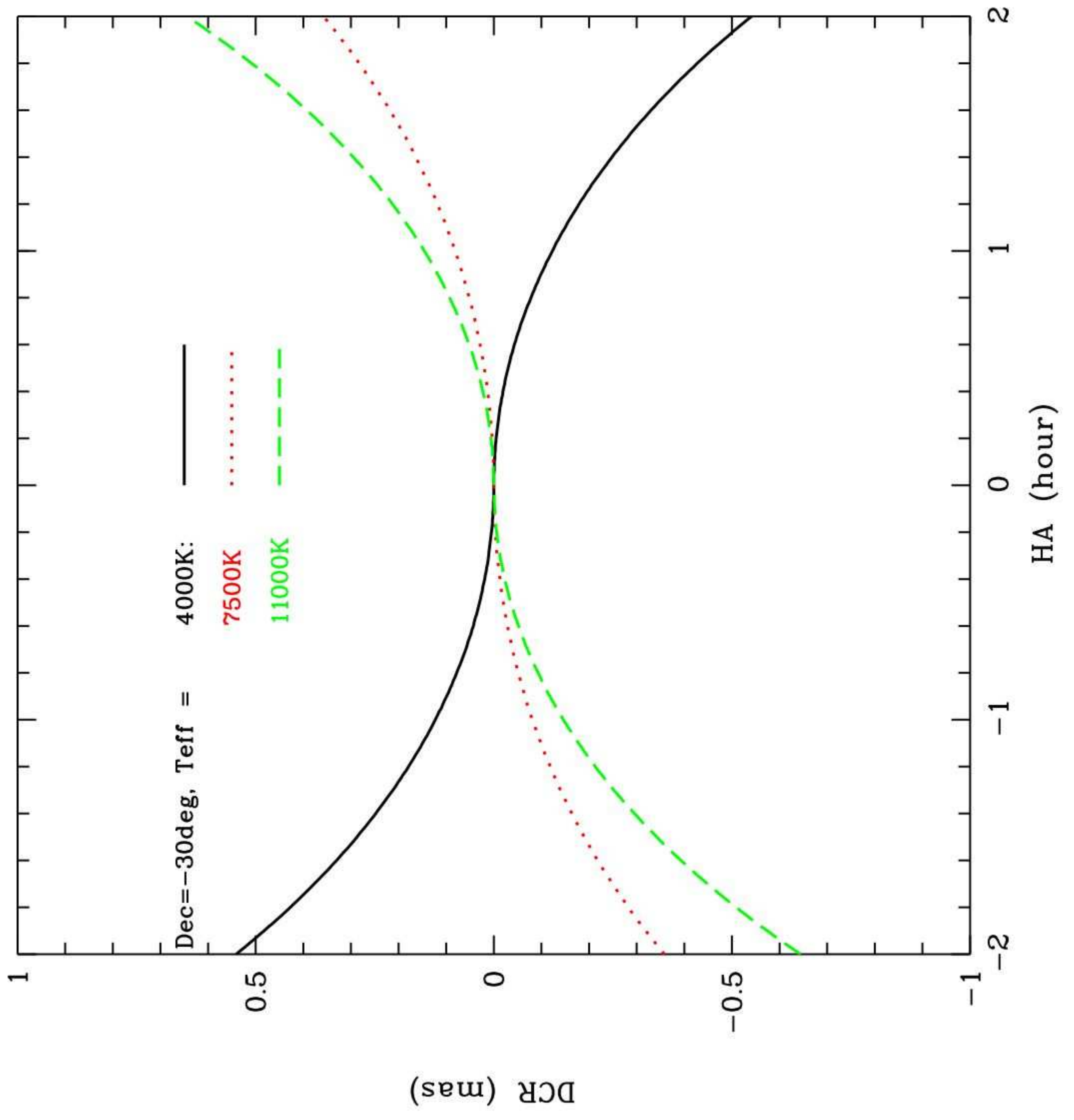}
\caption{\label{dcr}DCR effects in RA between a white dwarf of temperature 
T$_{\rm eff}$ and a mean background stars (T$_{\rm eff}$=5000K) at a
Declination of $-30\degr$ (representative of our sample) for various hour angles 
of observation. The DCR effects appear to be always lower than 0.5 mas for 
observations performed at less than 1 hour from meridian which is the case of the 
present project. The DCR effects are then negligible compared with other sources of 
astrometric error and were not taken into account in this work. }
\end{center}
\end{figure}

We present in Fig.~\ref{dcr} the effects of DCR in RA for 
white dwarfs situated at $\delta = -30\degr$, covering the range of temperatures 
of our targets. Fig.~\ref{dcr} demonstrates that the impact of 
DCR effects were always
less than 0.5~mas for observations taken with an hour angle of less than one 
hour. Therefore, our observations were made specifically so that the hour angle 
never exceeded one hour, and DCR corrections were not applied in this work.

\subsection{Impact of Pixel Scale Errors on Parallax}
Proper motions ($\mu_x,\mu_y$) and trigonometric parallax ($\pi_{xy}$) of targets 
are determined by comparing the ($x,y$) measurements expressed in pixels. A scaling 
factor $S_{\rm f}$, the image pixel scale, is applied to $\pi_{xy}$ to convert pixel 
measurements into physical units: $\pi = S_{\rm f}\pi_{xy}$; $d({\rm pc}) = \pi^{-1}$, 
with $S_{\rm f}$ expressed in $\arcsec$/pixel. 

Derivation of the pixel scale can be achieved through a cross--correlation between 
the ($x,y$) positions of stars on a given master frame to corresponding values of
($\alpha ,\delta$) for the subset of stars that are also in a reference catalogue. 
Here we used the 2MASS catalogue (\cite{cutr03}) to determine the orientation of the 
master frame on the sky and for the pixel scale determination. 
We selected the 2MASS catalogue as a reference catalogue because of its accuracy 
and density although we note the absence of proper motion corrections. 
Nevertheless the epoch difference between our observations and the 2MASS 
catalogue (3 years) would result in negligible corrections to the catalogue 
positions with respect to the catalogue errors.

Errors on the scale so determined, resulting from catalogue random errors, will 
produce errors in the distance determination of the target. It is therefore important 
to quantify the impact of the catalogue errors onto the distance of the target.

To measure this impact in the present work, we assumed $N$ reference stars 
equally spread over a square detector of side $A$. The classical equation 
relating the ($x,y$) measurements of a stars on the frame to its standard 
coordinates $X(\alpha,\delta),Y(\alpha,\delta)$ in the tangent plane to the 
celestial sphere is (with a similar equation in the $Y$ coordinate) 
\begin{equation}
X = (ax + by +c)1/F,
\end{equation}
where (a,b,c) are the unknown ``plate'' constants and~$F$ the focal length of 
the telescope (typically the value indicated in the reference manual).
$F$ is expressed in the same units as (x,y) and A (pixel, mm). 
It is then easy to show that a fair approximation of the variance of the 
estimation of parameter ($a$) is given by
\begin{equation}
\sigma_{a} \sim \sqrt{\frac{12}{N}} \frac{F}{A} \epsilon_{\rm cat} ,
\end{equation}
where $\epsilon_{cat}$ is the catalogue precision (expressed in radians). 
Similar results can be found in \cite{eich63}.
We can express the parallax (in radians) as:
\begin{equation}
\pi~=~\frac{a}{F}\pi_{xy},
\end{equation}
with
\begin{equation}
\sigma_{\pi}^2~=~\pi_{xy}^2~\frac{1}{F^2} \sigma_a^2,\\
\sigma_{\pi}~=~\frac{\pi_{xy}}{F}\sqrt{\frac{12}{N}}\frac{F}{A}\epsilon_{cat}
\end{equation}
with $F\sim13$m, $A\sim0.03$m, we evaluate here $\sigma_{\pi}~\sim~10^{-4}\pi$.
The impact of the error of the catalogue on the parallax of the target is far  
below the measurement errors (typically a few milliarcseconds) and are therefore 
negligible.

\subsection{Global Solution: Relative Parallax}
The astrometric reduction of the whole set of data of each field is performed 
iteratively through a global central overlap procedure (\cite{hawk98}, \cite{eich97}) 
in order to determine simultaneously the position, the proper motion and the parallax 
of each object of the field.\\

The following condition equations are written for each star on each of
the N frames considered (including the master frame). 
These equations relate the measured coordinates to the stellar astrometric 
parameters: 

\begin{equation}
X_{0} +	\Delta X_{0} + \mu_{X} (t-t_{0}) + \pi F_{X}(t) = a_{1} x(t) + a_{2} y(t) + a_{3} 
\end{equation}
\begin{equation}
Y_{0} +	\Delta Y_{0} + \mu_{Y} (t-t_{0}) + \pi F_{Y}(t) = b_{1} x(t) + b_{2} y(t) + b_{3} 
\end{equation}

\noindent where ($X_{0}, Y_{0}$) are the known standard coordinate of the star at the 
epoch $t_0$ of the master frame, and ($x(t), y(t)$) its measured coordinates
on the frame (epoch $t$) to be transformed into the master frame system.
$\Delta X_{0}$, $\Delta Y_{0}$, $\mu_{X}$, $\mu_{Y}$ and $\pi$ are 
the unknown stellar astrometric parameters: ($\Delta X_{0}$, $\Delta Y_{0}$) yield
correction
of the standard coordinates of the star on the master frame, ($\mu_{X}$, $\mu_{Y}$) 
are the projected proper
motion in RA $*cos(\delta)$ and Dec, and $\pi$ is the parallax. 
Coefficients ($a_{i}$, $b_i$) are
the unknown frame parameters which describe the 
transformation to the master frame system.
$(F_{X},F_{Y})$ are the parallax factors in standard 
coordinates.
The unknowns of this large over--determined system of equations are the stellar 
astrometric parameters of each object, and the transformation coefficients of each 
of the $N$ frames
considered. The system of equations is singular and therefore the
derived solution is not unique; any solution will depend on the starting 
point of the iterations. The usual technique to obtain a particular solution 
is to introduce a set of constraints that the solution must satisfy. In this 
work we chose to set strictly to zero the mean parallax of the reference 
stars. 

We used a Gauss--Seidel type iterative method to solve the set of
equations.  At the first iteration all stellar parameters are assumed null, 
we then computed the plate constants
which are injected into the system of equations to derive the stellar
parameters. These results are then used as the starting point of the
following iteration. The iterative procedure converges usually at the second
or third iteration.  A test of elimination at $3\sigma$ is
applied to remove poor observations either in the master frame 
fit or in the stellar parameters fit. The stellar parameters fit equations 
have been weighted by the mean residual of the master frame fit. 
This weighting represents the quality of the measurements. The stars used 
for the master frame fit are called here reference stars.

We applied this global treatment to the various observations of the 15 fields 
observed and we derived for the targets a proper motion and parallax with 
associated variances.

\subsection{Conversion from Relative to Absolute Parallax}
The parallaxes that we derived for our targets are relative to the reference 
stars (for which we used the constraint $\sum{\pi} = 0$), supposed placed 
at infinite distance. In fact these reference stars are at a finite distance 
from Sun. We must therefore correct  the relative parallax of the target from 
an estimate of the mean distance of the reference stars to obtain the 
absolute parallax of the target. The choice we made to keep as many reference 
stars as possible in our calculation is interesting because statistically faint 
stars have smaller parallax and require smaller correction.   

There are several ways to estimate the mean distance of reference stars:
statistical methods relying on a model of the Galaxy; spectroscopic parallax;
and photometric parallax. For the corrections from relative parallax to 
absolute parallax we used a statistical method relying on simulations using 
the Besan\c con Galaxy model (\cite{robi94}) to derive the theoretical mean 
distance of reference stars. A simulation of each observed field was performed,
providing catalogues of distance and apparent magnitude of simulated stars. 
We computed in these catalogues mean distances and associated dispersion in 
magnitude bins of 0.2 mag, establishing a table of theoretical distances with 
respect to apparent magnitude. Then we considered our observed fields and we 
computed the weighted mean parallax and associated dispersion of our reference 
stars using the theoretical table. Finally we added  this mean parallax of 
reference stars to the relative parallax of our target leading to the absolute 
parallax of the white dwarfs. 

We give in Table \ref{corrpi}  the relative to absolute corrections in 
milliarcseconds as found from the Besan\c con Galaxy model in each of the field 
treated.

\begin{table}[h]
\caption{
\label{corrpi}
Relative to absolute corrections $\Delta \pi$ and associated RMS ($\sigma$) as found 
from the Besan\c con Galaxy model in the Galactic direction (l,b) together with 
number of reference stars (N*) in magnitude interval [Jmin,Jmax].}
\begin{center}
\begin{tabular}{lrrrrrrr}
\hline
   &   &  &  &  & &   &    \\
 Target	     &       	   l   &    b   &  $\Delta \pi$ & $\sigma$& N* &  $J_{min}$ & $J_{max}$  \\
 	             & [$\degr$]  & [$\degr$]  & \multicolumn{2}{c}{[mas]} &    &  \multicolumn{2}{c}{[mag]}      \\
\hline
\hline
   &   &  &  &  & &   &    \\
 WD2214-390  &    2.79 & -55.37 &  1.3 & 0.3& 38 & 13.1 & 16.2 \\
 WD2242-197  &   40.01 & -59.42 &  1.0 & 0.3& 97 & 14.0 & 18.4 \\
 WD2259-465  &  344.30 & -60.62 &  1.1 & 0.2& 83 & 13.6 & 18.0 \\
 LHS542      &   72.40 & -59.70 &  1.2 & 0.3& 42 & 13.4 & 17.0 \\
 WD2324-595  &  321.83 & -54.34 &  1.1 & 0.2& 62 & 13.3 & 17.0 \\
 WD2326-272  &   27.66 & -71.06 &  1.3 & 0.4& 80 & 14.2 & 18.7 \\
 LHS4033     &   90.24 & -61.96 &  1.3 & 0.2& 39 & 14.2 & 16.5 \\
 LHS4041     &  351.44 & -74.66 &  1.4 & 0.3& 37 & 13.5 & 16.2 \\
 LHS4042     &    6.55 & -76.61 &  1.5 & 0.4& 38 & 13.3 & 16.6 \\
 WD0045-061  &  118.54 & -68.96 &  1.5 & 0.3& 54 & 13.5 & 17.7 \\
 F351-50     &  314.26 & -83.50 &  0.3 & 0.2& 53 & 14.1 & 18.1 \\
 LP586-51    &  128.88 & -63.30 &  1.3 & 0.3& 47 & 14.1 & 17.4 \\ 
 WD0135-039  &  149.30 & -64.53 &  1.3 & 0.2& 82 & 14.4 & 19.0 \\
 LP588-37    &  150.44 & -61.52 &  1.4 & 0.2& 57 & 13.6 & 17.7 \\
 LHS147      &  178.72 & -73.56 &  1.5 & 0.3& 43 & 13.4 & 16.8 \\
   &   &  &  &  & &   &    \\
\hline
\end{tabular}
\end{center}
\end{table}

\section{Results}
\label{pis}
\subsection{Distances of Halo White Dwarf Candidates}
We present in Table \ref{pi} the proper motions and absolute parallaxes of the 
fifteen halo white dwarf candidates as derived from this work together with 
their absolute magnitude $M_{\rm V}$ computed using 
CCD $V$ magnitudes from Bergeron et al.~(2005).

\begin{table*}
\caption{
\label{pi}
 Proper motion and absolute parallaxes of the fifteen halo white dwarf
candidates, where $\mu_{\alpha*} = \mu_{\alpha}cos(\delta)$ and  
$\sigma_{\mu}$=$\sigma_{\mu_{\alpha*}}$=$\sigma_{\mu_{\delta}}$; $\pi$  
and $\sigma_{\pi}$ are the parallax and its precision, Dist the derived 
distance in parsec and $M_{\rm V}$ the absolute magnitude. No value is given 
for Dist and Mv when the parallax is not better than 3 $\sigma$. 
N* is the number of reference 
stars and Nf the Number of frames. $D_{phot}$ is the photometric distance from 
OHDHS and V is extracted from Bergeron et
al.~(2005) when available, otherwise (cases marked by an asterix) it comes 
from Salim et al.~(2004). Note that LHS~4041 is in the OHDHS sample, but
is not listed in OHDHS Table~1 (see Table~4 of Salim et al.~2004)}
\begin{center}
\begin{tabular}{l c c c r r r c c c r c c c r r}
\hline
   &   &  &  &  & &   &  &  & &\\
Name     & $\alpha$ & $\delta$ & Epoch & V &  $\mu_{\alpha*}$  &    $\mu_{\delta}$ & $\sigma_{\mu}$  &   $\pi$  &  $\sigma_{\pi}$   & Dist         &$M_{v}$& N* &Nf & $D_{phot}$ &\\
	     & \multicolumn{2}{c}{[J2000]}&[yr] &[mag]    & \multicolumn{3}{c}{[mas/yr] }                                                    &   \multicolumn{2}{c}{[mas]}& [pc]  &  [mag]  &     &    &[pc]  & \\
   &   &  & &  &  & &   &  &  & & \\
	     \hline
  	     \hline
   &   &  & &  &  & &   &  &  & & \\
WD2214--390  &22 14 34.727 & --38 59 07.05  & 2003.5 & 15.92          &  1009 &  --350 & 2.9 & 53.5 & 2.6 &  19    &  14.78  & 38 & 28 &24 &\\
WD2242--197  &22 41 44.252 & --19 40 41.41  & 2003.5& 19.74          &   359 &   +48   & 3.1 & 11.1 & 2.3 &  90     &  14.89  & 97 & 27 &117 &\\
WD2259--465  &22 59 06.633 & --46 27 58.86  & 2002.9& 19.56          &   402 &  --153  & 1.8 & 22.7 & 1.3 &  44     &  16.49  & 83 & 32 &71 &\\
LHS542            &23 19 09.518 & --06 12 49.92  & 2003.5& 18.15         &  --615 & --1576 & 1.8 & 29.6 & 1.8 &  34     &  15.58  & 42 & 33 &42 &\\
WD2324--595  &23 24 10.165 & --59 28 07.95  & 2003.5& 16.79          &   136 &  --562   & 1.8 & (3.1) & 1.5 & ---- & ---- & 62 & 25&58 & \\
WD2326--272  &23 26 10.718 & --27 14 46.68  & 2002.9& $^{*}19.92$&   574 &   --85    & 2.7 & (6.2) & 2.4 & ----   & ---- & 80 & 17&108 &\\
LHS4033          &23 52 31.941 & --02 53 11.76  & 2002.9& 16.98         &   631 &   298    & 2.5 & 30.1 & 1.8 &  33    &  14.38  & 39 & 26 &63 &\\
LHS4041         &23 54 18.793 & --36 33 54.60  & 2002.9& $^{*}15.46$&    21 &  --662    & 1.8 & 13.4 & 1.5 &  75    &  11.10  & 37 & 27 &59 &\\
LHS4042         &23 54 35.034 & --32 21 19.44  & 2003.5& 17.41         &   421 &   --37    & 2.2 & 13.9 & 1.8 &  72    &  13.13  & 38 & 25 &85 &\\
WD0045--061  &00 45 06.325 & --06 08 19.65  & 2002.9& 18.26         &   111 &  --668   & 1.9 & 30.1 & 1.9 &  33    &  15.59  & 54 & 27 &44 &\\
F351--50         &00 45 19.695 & --33 29 29.46  & 2003.5& 19.01         &  1820 & --1476 & 2.1 & 28.3 & 1.4 &  35     &  16.63  & 53 & 34 &37 &\\
LP586--51       &01 02 07.181 & --00 33 01.82  & 2002.9& 18.18         &   350 &  --118   & 3.6 & (2.4) & 2.7 & ----    & ----  & 47 & 24 &120 &\\
WD0135--039  &01 35 33.685 & --03 57 17.90  & 2002.9& 19.68         &   456 &  --180   & 3.4 & 13.3 & 2.9 &  75    &  15.26  & 82 & 21 &146 &\\
LP588--37       &01 42 20.770 & --01 23 51.38  & 2002.9& $^{*}18.50$&   112 &  --328   & 3.4 & (1.4) & 4.5 & ----    &  ---- & 57 & 17 &120 &\\
LHS147           &01 48 09.120 & --17 12 14.08  & 2002.9& 17.62         &  --115 & --1094 & 2.1 & 14.8 & 1.8 &  68     &  13.46  & 43 & 29 &71 &\\
\hline
\end{tabular}
\end{center}
\end{table*}

One notices that WD2326--272, LP586--51, LP588--37, and WD2324--595 are too 
distant to have a measurable parallax. Eleven objects are at 
distances ranging from 19 pc to 90 pc from the Sun. The parallax errors are about 
1--2 mas corresponding to relative precisions of 5 to 20\%. WD2214--390, which 
is the closest and brightest object, has a $ \sigma_{ \pi}$ = 2.6 mas. 
This poor precision is due to the short 
exposure time used to avoid saturation problems and corresponding 
lower signal--to--noise ratio. 

We present in Figs~\ref{pifig1} and~\ref{pifig2} the positions (empty circles), 
their weighted mean (filled circles) and associated error bars  at each epoch of 
observation, together with the fitted path for the eleven most significant 
parallaxes, where $\pi / \sigma_{ \pi} \ge 4$.

\subsection{Comparison with Published Distances}
\begin{figure}[ht]
\begin{center}
\includegraphics*[width=8cm,angle=-90]{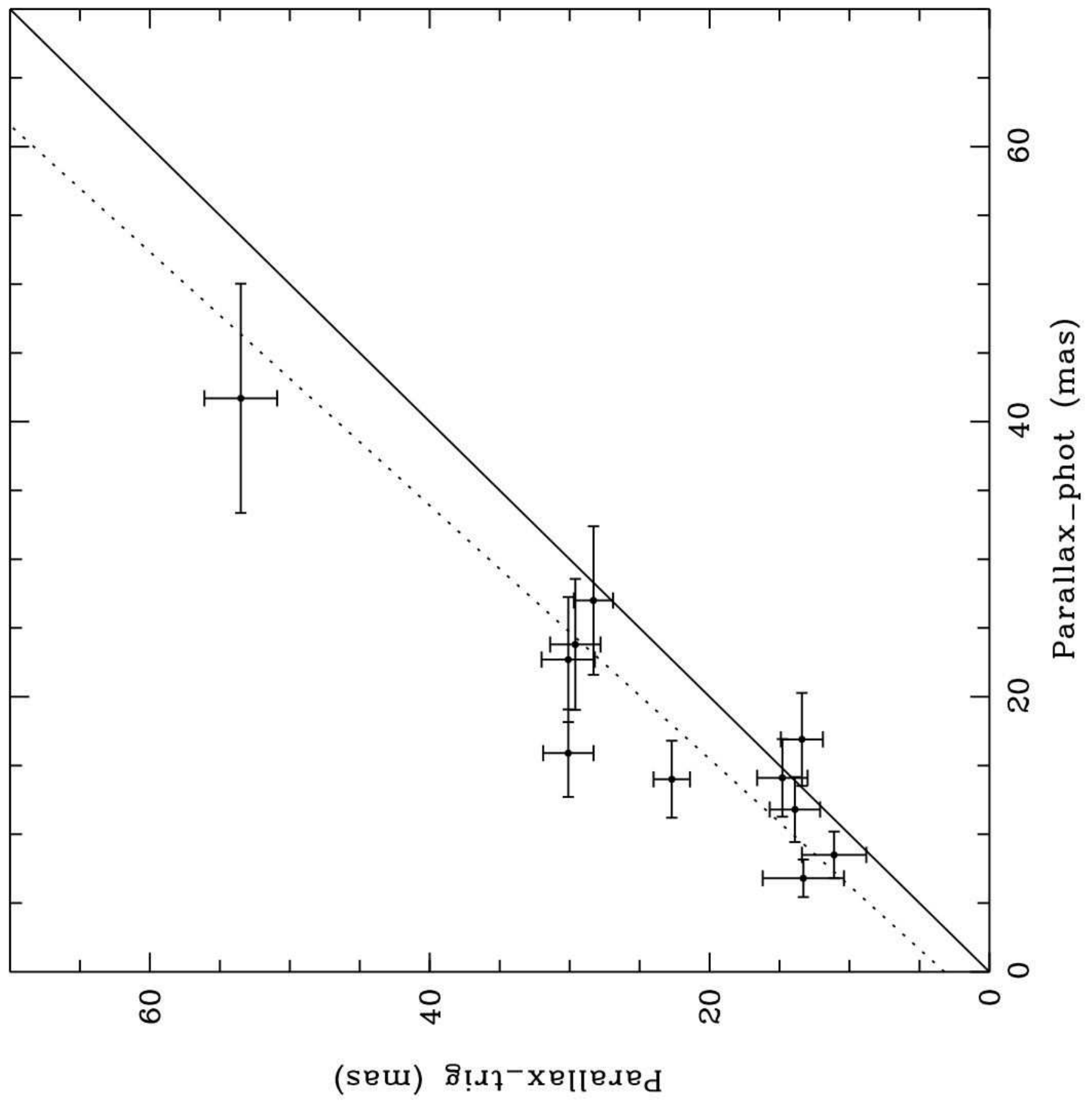}
\caption{\label{Ddphot} Comparison of parallaxes derived in this work with 
photometric parallaxes from OHDHS (errors are assumed 20\% for
$\pi_{phot}$). Parameters of a weighted linear 
regression (diagonal line) between both types of parallaxes are  
$\pi = 1.08 \pi_{phot} + 3.21$ [mas] 
with a reduced $\chi^{2} = 8.06$. The photometric distances are 
systematically larger than the astrometric ones.}
\end{center}
\end{figure}

We have compared our results with available data from the literature, 
employing both trigonometric and photometric parallaxes measured previously. 
We give in Table~\ref{dpi} the comparison with published trigonometric 
parallaxes and in Figure~\ref{Ddphot} a comparison of the parallaxes derived 
in this work with photometric parallaxes (from OHDHS, where photometric 
parallax errors were 20\%). Parameters 
of a weighted linear fit between photometric and trigonometric parallaxes 
are:~$\pi_{trig} = a.\pi_{phot}+b$ 
with a~=~1.08+/-0.08 and b~=~3.21+/-1.56 [mas] with a reduced $\chi^{2}$~=8.06. 
\begin{table}[h]
\caption{\label{dpi}Comparison of trigonometric parallaxes from this work 
($\pi_{\rm This work}$) with published data ($\pi_{\rm ext}$)
for LHS 147 (\cite{vana95}), LHS 4033 (\cite{dahn04}) and 
LHS 542 (\cite{berg05}).}
\begin{center}
\begin{tabular}{lccr}
\hline
 Target	     & $\pi_{\rm This work} $  &  $ \pi_{\rm ext} $  &  $\Delta \pi$ \\
 	             & [mas]               & [mas]           & [mas]     \\
\hline
\hline
LHS 542        &  29.6 +/- 1.8    &  32.2 +/- 3.7  & 2.6\\
LHS 147        &  14.8 +/- 1.8    &  14.0 +/- 9.2 & --0.8\\
LHS 4033      &  30.1 +/-1.8     &  33.9 +/- 0.6 & 3.8 \\
\hline
\end{tabular}
\end{center}
\end{table}

Our parallaxes are in excellent agreement with the 3 previously published 
trigonometric parallaxes, within the errors (which are considerably smaller
in two cases than published values). In Fig.~\ref{Ddphot} one notices a clear 
systematic tendency of photometric parallaxes to be underestimated. This 
overestimation of OHDHS distances is of importance in the calculation of WD 
kinematics and space density.

\subsection{Proper Motions}
\begin{figure}[h]
\begin{center}
\includegraphics*[width=8cm,angle=-90]{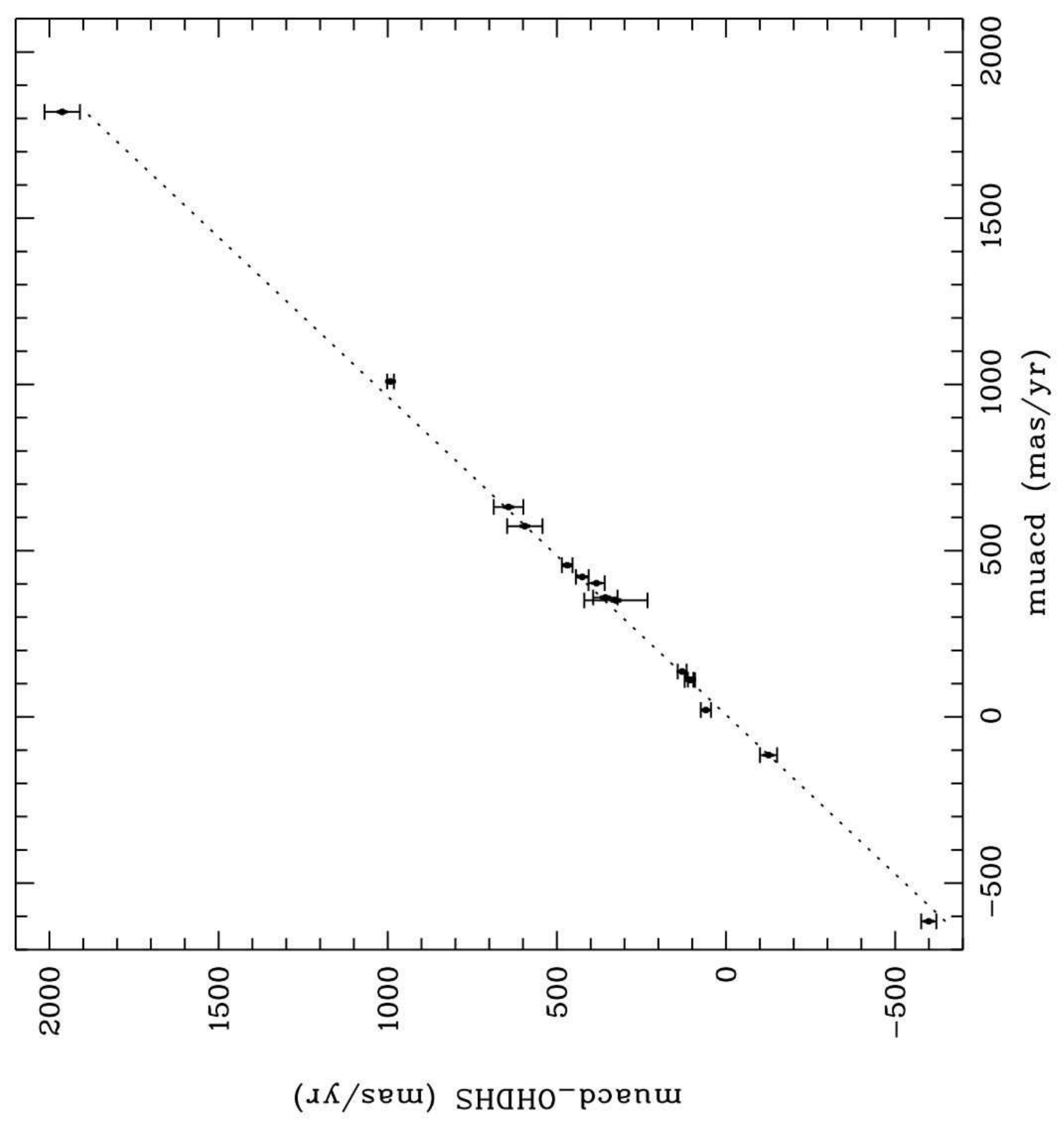}
\caption{\label{muacd}
Comparison of proper motions in RA cos($\delta$) with the OHDHS proper motions. 
Error bars are drawn in both coordinates but since the present work has much 
higher precision than the photographic astrometry, error bars in abscissae
are not visible. 
The slope of a linear regression (dotted line) is 1.04 $\pm$ 0.02 indicating 
good accordance between both proper motion data sets with a reduced 
$\chi^{2} = 3.7$. }
\end{center}
\end{figure}

\begin{figure}[h]
\begin{center}
\includegraphics*[width=8cm,angle=-90]{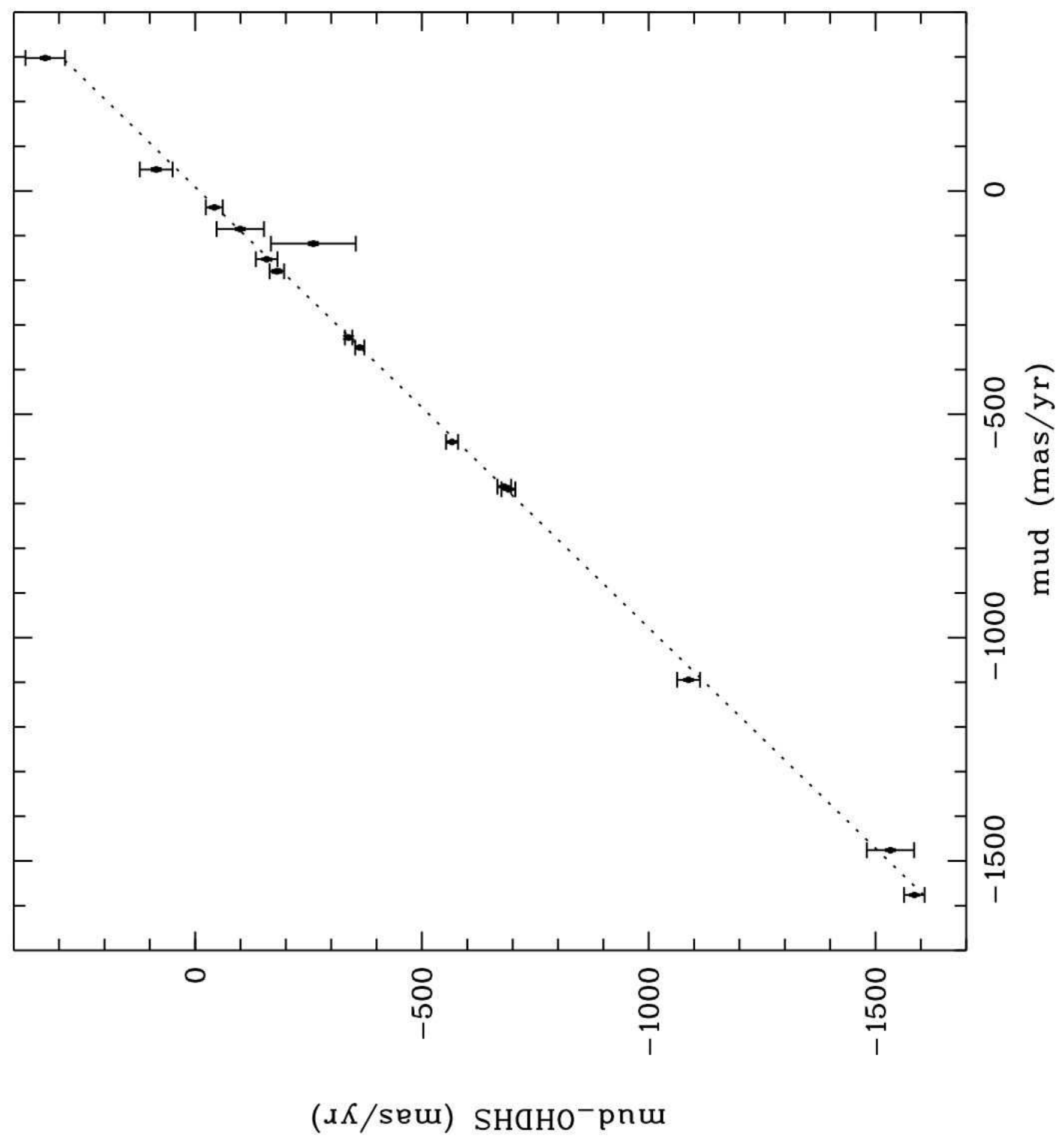}
\caption{\label{mud}
Comparison of proper motions in Declination derived in this work 
with the OHDHS proper motions. Error bars are drawn in both coordinates but since 
the present work has much higher precision than the 
photographic astrometry, error bars in abscissae are not visible. The slope of 
a linear regression (dotted line) is 1.01 $\pm$ 0.02 indicating a good accordance 
between both proper motion data sets with a reduced $\chi^{2} = 0.7$.}
\end{center}
\end{figure}

We have compared the proper motions derived here with the OHDHS proper motions 
in order to check wether some systematic effects could affect our proper motions 
derived on a 1.5~yr time span and, as a result, our parallaxes. 
We present this comparison in Fig. \ref{muacd} and Fig. \ref{mud}. 
Error bars are drawn in both coordinates but since the present work has much higher 
precision than the photographic astrometry, the error bars in x are not visible. 
The slope of a linear regression between proper motions in $\alpha$ cos($\delta$) 
derived in this work with the OHDHS proper motions is 1.04 $\pm$ 0.02 with a 
reduced $\chi^{2} = 3.7$. The equivalent linear fit  in proper motions in Declination 
has a slope of 1.01 $\pm$ 0.02  with a reduced $\chi^{2} = 0.7$. 
For F351-50 (the largest error bars in both figures), the accordance in RA and Dec 
proper motions is not good. This is due to a known problem of contamination by a 
background galaxy of the Schmidt plate measurements used in the
OHDHS work. Nevertheless the accordance is within 2$\sigma$.
These comparisons show excellent agreement 
between both sets of proper motions, and argue against any systematic effects 
from the present work. 


\subsection{Space Velocities}
We derived the Galactic space velocities U, V, W (\cite{john87})  
for the white dwarfs using the distances and proper motions 
measured here together with radial velocities  from \cite{sali04} (data 
available for 9 of the 15 white dwarfs treated here). Salim's observed 
radial velocities were corrected for a mean gravitational redshift of 
+28km/s as suggested by the authors in their paper except in the case of the 
very massive white dwarf LHS4033 were the correction was taken from \cite{dahn04}. 
U is radial toward 
the Galactic center, V is in the direction of rotation and W perpendicular 
to the Galactic disk. U,V and W were corrected for the Sun's peculiar velocity 
(\cite{miha81}). When no radial velocity was available from other studies, we 
assumed $V_r = 0$ km/s. This approximation is acceptable due to its minor impact 
on U,V velocities since the targets are located close to South Galactic Cap
(the effect was investigated in OHDHS and shown to be negligible).

\begin{figure}[h]
\begin{center}
\includegraphics*[width=8cm,angle=-90]{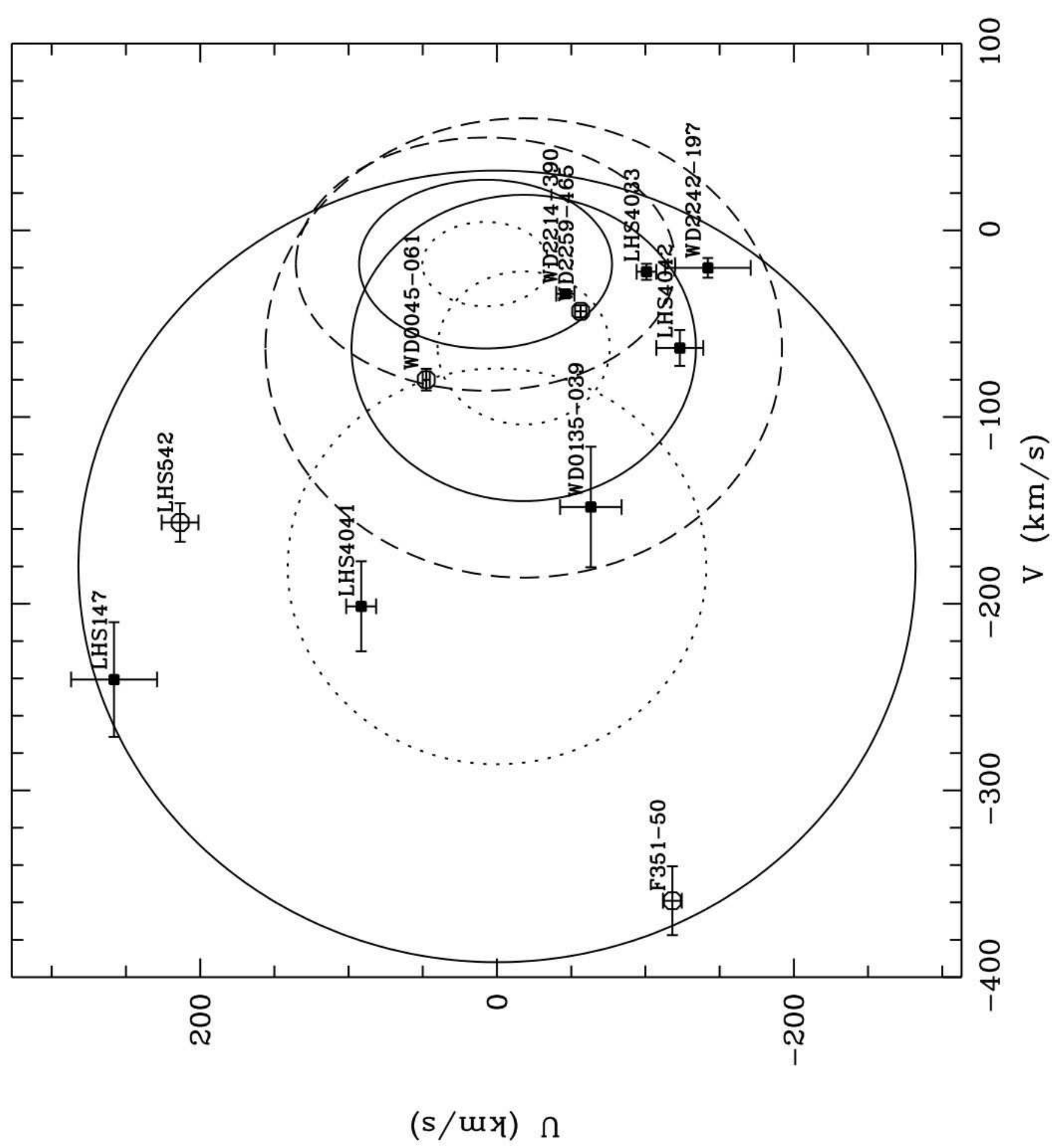}
\caption{\label{U_V}
Distribution of velocities in the Galactic plane together with the velocity 
dispersion for the disk (right most)(1, 2 and 3 $\sigma$), thick disk 
(middle)(1, 2 and 3 $\sigma$)( \cite{fuhr04}) and halo (left) (1 and 2$\sigma$) 
(\cite{chib00}). 
Filled squares correspond to objects with a measured radial velocity (\cite{sali04}) 
while open circles correspond to objects with no $V_r$ measurement. 
Only objects with  parallax measured at the $4\sigma$ level or better are plotted. }
\end{center}
\end{figure}

\begin{figure}[h]
\begin{center}
\includegraphics*[width=8cm,angle=-90]{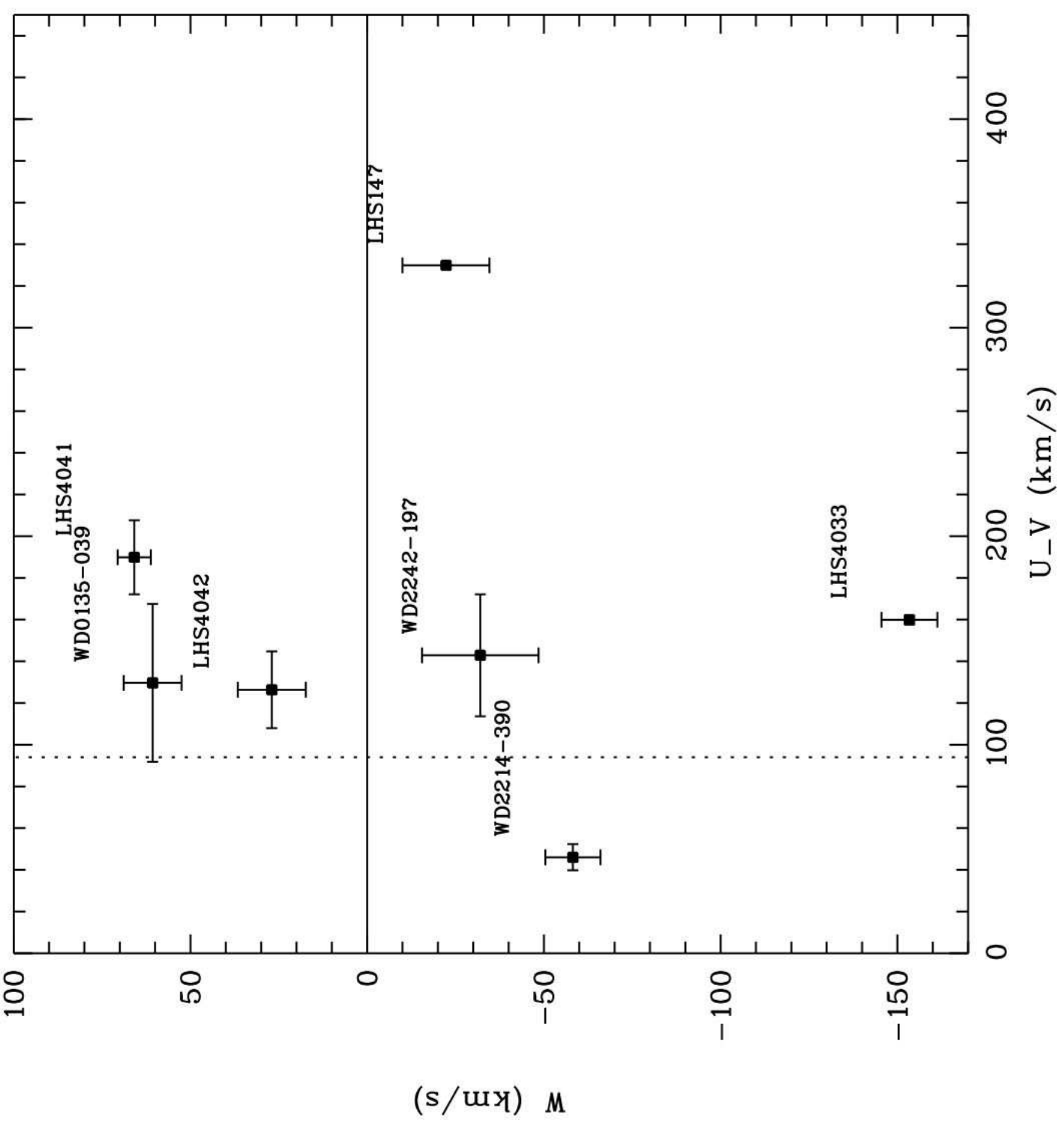}
\caption{\label{U_V_W}
Component of motion perpendicular to the Galactic plane (W) as function of 
$\sqrt{U^{2}+V^{2}}$. 
Only objects with parallax measured at the $4\sigma$ level or better and with 
available radial velocity (\cite{sali04}) arre plotted. 
The vertical line is the OHDHS $\sqrt{U^{2}+V^{2}}$ = 94 km/s cut.  }
\end{center}
\end{figure}

We present in Figure \ref{U_V}  the distribution of velocities in the Galactic plane 
together with the velocity 
dispersion for the disk (right most)(1, 2 and 3 $\sigma$), thick disk 
(middle)(1, 2 and 3 $\sigma$)( \cite{fuhr04}) and halo (left) (1 and 2$\sigma$) 
(\cite{chib00}) and in Figure 
\ref{U_V_W} the component of motion perpendicular to the Galactic plane. 
These two figures concern the 11 objects with  parallax measured at the $4\sigma$ 
level or better.

In Fig.~\ref{U_V} one notices that 4 of the 11 studied WDs have a velocity 
incompatible at the $3\sigma$ level with the kinematic of the disk and of the thick 
disk and that 6 of them are incompatible at a $2\sigma$ level. No star lies within the 
$1\sigma$ ellipse of the disk, primarily because of selection effects in the 
original proper motion survey that OHDHS based is based upon \cite{hamb05}. 

Obviously the choice of the center and dispersions of halo, thick disk and 
disk ellipses is
critical to classify objects as belonging to a particular population. 
We adopted recent values which are in in the range of the values cited by 
\cite{reid05} in his review: Disk (\cite{fuhr04})~: (U,V)~=~(7.7, $-$18.1) km/s, 
($\sigma_{U},\sigma_{V}$) = (42.6, 22.6) km/s;
 thick disk (\cite{fuhr04}): (U, V) = (-18, $-$63) km/s, 
($\sigma_{U},\sigma_{V}$) = (58, 41) km/s; 
halo (\cite{chib00}): (U,V) = (0, $-$180) km/s, 
($\sigma_{U},\sigma_{V}$) = (141, 106) km/s.

\section{Discussion}
\label{dis}
As discussed above, OHDHS sparked a lively debate about whether stellar 
remnants contribute to a significant fraction of the baryonic component of
the putative dark matter halo of our Galaxy. The main criticisms have 
concerned interpretation, and we do not address those here. However, the
photographic photometry and use of a single photometric parallax relation
are also potential sources of systematic error. Both Salim et al.~(2004)
and Bergeron et al.~(2005) have shown that the original photometry presented
in OHDHS was as accurate as could be expected. Here, we address the question
of the accuracy of photometric parallaxes directly via trigonometric
determination of distances.

 In Fig.~\ref{Ddphot} we compare the trigonometric parallaxes derived here with 
 the OHDHS photometric parallaxes. Parameters of a weighted linear 
regression between both types of parallaxes are  $\pi$ = 1.08 $\pi_{phot} + 3.21$ 
with a reduced $\chi^{2}$ = 8.06. A clear underestimation of photometric parallaxes 
is visible in this figure with only one point below the diagonal 
and three points more than $3\sigma$ above the relation. 
With the usual caveat of small number statistics,
this indicates some level of non--Gaussian scatter, or at least a mean value 
for the relation that is not coincident with $\pi = \pi_{phot}$. 
The photometric parallax overestimates the distance. This leads, of course, to an
overestimation of tangential space velocities based on proper motion and distance
(as an aside, we note that the quoted photometric 
parallax errors of 20\% were conservatively overestimated by OHDHS). 

It is interesting to note that the mass distribution of hot 
($T_{\rm eff} > 12,000$ K) DA WDs is not Gaussian and has a broad tail on 
the high mass side
(\cite{made04}). Given that radius $r \propto m^{-1/3}$ for WDs, we would
expect photometric parallaxes to tend to overestimate rather than 
underestimate distances since some of the sample may have higher than 
average mass, and correspondingly smaller 
radii, placing them nearer to the
Sun than typical objects of the same colour. Adding in a sprinkling of
higher mass WDs with helium--dominated atmospheres will introduce further
systematic overestimation of distances. It is almost certainly the case
that the discrepant photometric parallaxes for WD2259--465 and WD0135--039
are caused by these effects; indeed, this has been shown to be the case
for LHS~4033 which has a mass $m\sim1.3$~M$_{\odot}$ (\cite{dahn04}).
On the other hand, the low--mass side of the mass distribution is
by no means perfectly Gaussian
(e.g.~due to low-mass, helium--core white dwarfs formed in close binaries).
Moreover, any overestimation in distance leads to a corresponding underestimate of
space density using the 1/Vmax technique. So the interpretation of the
results from this relatively small sub--sample is rather complicated, and 
it is only through detailed simulations compared with much larger samples
that significant progress is likely to be made concerning the question of the 
kinematic population of such objects.

\begin{figure}[h]
\begin{center}
\includegraphics*[width=8cm,angle=-90]{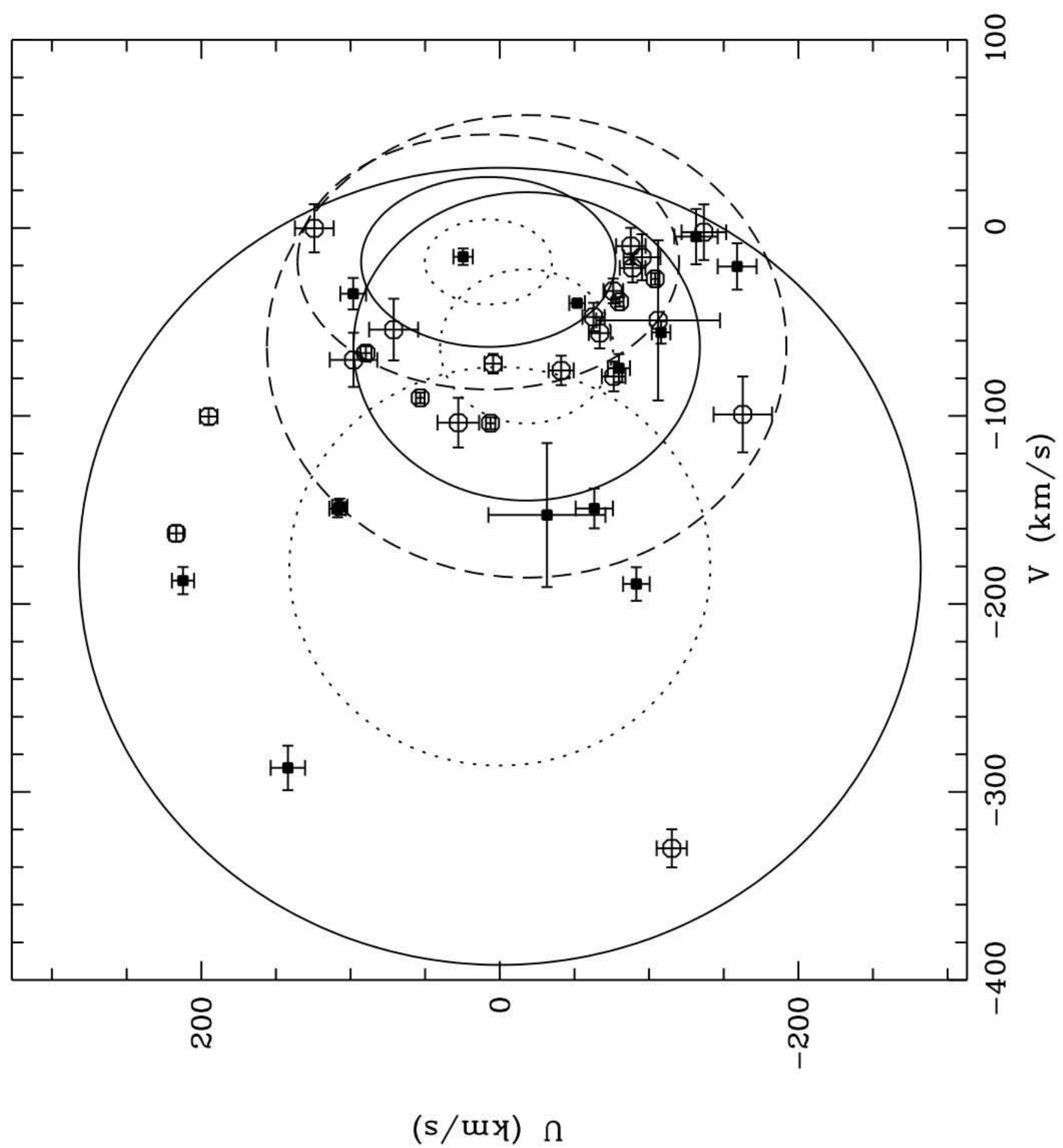}
\caption{\label{opp_fig} 
Distribution of velocities of the original OHDHS sample with recalibrated parallaxes 
in the Galactic plane together with the velocity dispersion 
for the disk (right most)(1, 2 and 3 $\sigma$), thick disk (middle)
(1, 2 and 3 $\sigma$)( \cite{fuhr04}) and halo (left) (1 and 2$\sigma$) 
(\cite{chib00}). Filled squares correspond to objects with a measured radial 
velocity (\cite{sali04}) while open circles correspond to objects with no 
$V_r$ measurement. }
\end{center}
\end{figure}

\begin{figure*}[ht]
\begin{center}
\includegraphics*[width=18cm,angle=0]{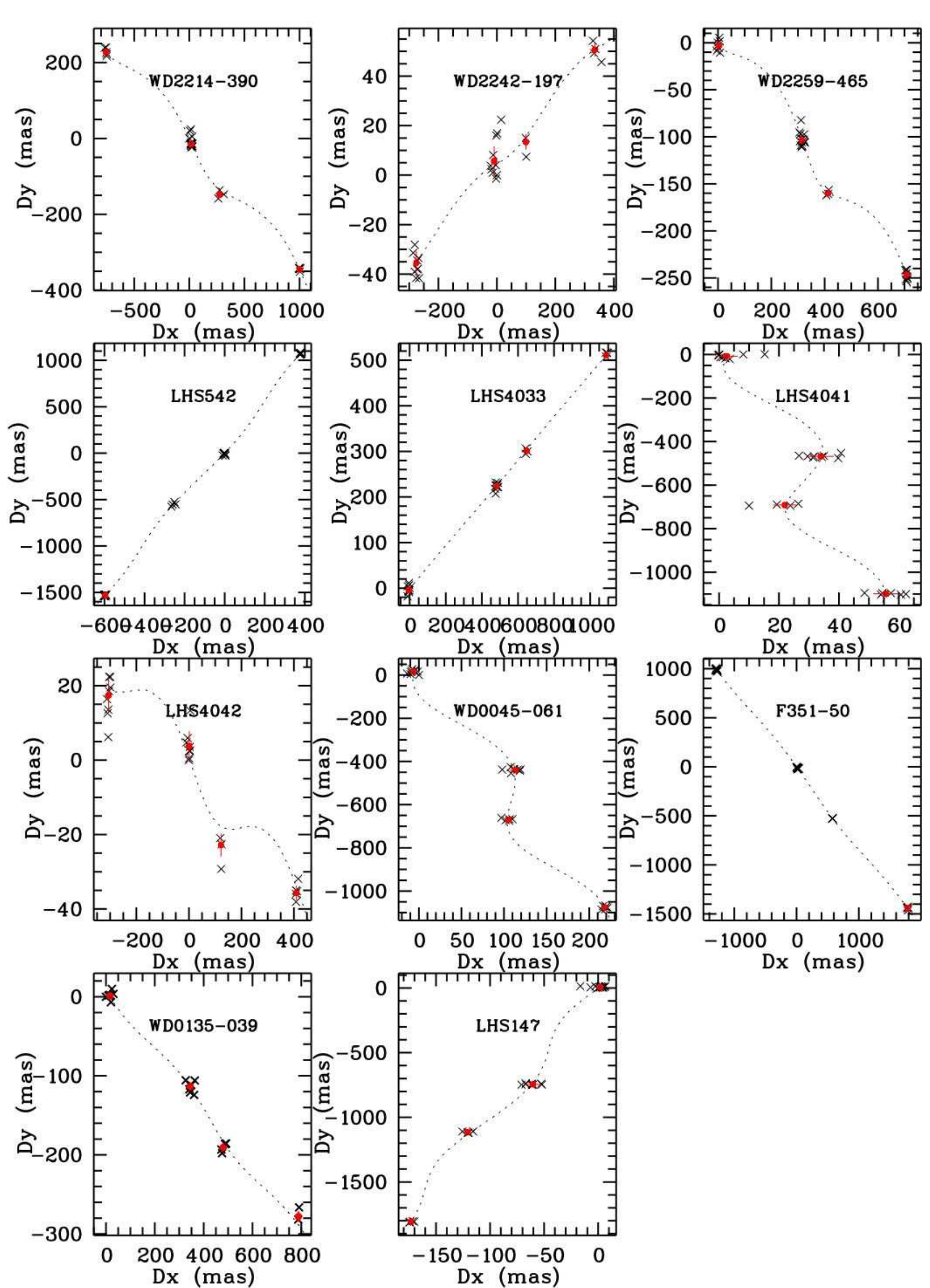}
\caption{\label{pifig1} Observations along the fitted path expressed in mas.}
\end{center}
\end{figure*}

From the comparison of trigonometric and photometric parallaxes (Fig. \ref{Ddphot}) 
we recalibrated photometric distances of the original OHDHS sample and, using 
radial velocities from \cite{sali04}, we derived their associated recalibrated 
space velocities. We present the recalibrated UV plane for the entire OHDHS sample 
in Fig. \ref{opp_fig}.

When compared to Fig. 3 of OHDHS, the number of halo objects has diminished. 
From the 38 original OHDHS halo candidates, 
16 appear compatible with a halo status based on a $2\sigma$ cut with 
the disk and thick disk velocity distributions (a $3\sigma$ cut would reduce 
this number to 7), the remaining objects being now located within the disk and 
thick disk 2 sigma ellipses.
In the literature there is a large spread of the proposed values to characterise 
the thick disk and halo populations in terms of kinematics. For instance in \cite{reid05} 
the velocity dispersions for thick disk vary from 50 to 69 km/s in the U direction and 
from 39 to 58 km/s in the V direction. Even the center of velocity ellipsoid varies
from --30 to --63 km/s in the $<V>$ coordinate from one author to another. All this makes it
very difficult to separate objects into halo and thick disk populations and requires a 
more detailed analysis which is beyond the scope of the present paper.

The conclusions of OHDHS about local halo WD density must be now reanalysed since 
the volume explored by their survey has changed (re-calibrated distances) and the 
number of halo candidates has also changed. This will be the subject of a forthcoming 
paper.

\section{Acknowledgements}
The authors wish to thank G.~Daigne for helpful comments and CAPES/COFECUB, FAPESP
organizations and INR for supporting the project.

\begin{thebibliography}{}

\bibitem[Alcock et al.~1999]{alco99}
Alcock, C. et al. 1999, in ASP Conf. Ser. 165, The third Stromlo Symposium: 
the Galactic Halo, ed. B.K. Gibson, T.S. Axelrod and M.E. Putman (San 
Francisco:ASP),362

\bibitem[Bergeron \& Leggett~2002]{berg02}
Bergeron, P., Leggett, S.~K.~2002, ApJ, 580, 1070

\bibitem[Bergeron~2003]{berg03}
Bergeron, P.~2003, ApJ, 586, 201

\bibitem[Bergeron et al.~2005]{berg05}
Bergeron, P.; Ruiz, M.--T.; Hamuy, M.; Leggett, S.~K.; Currie, M.~J.; Lajoie, 
C.--P.; Dufour, P.~2005, ApJ, 625, 838

\bibitem[Calchi~Novati et al.~2005]{calc05}
Calchi~Novati, S.~et al.~2005, A\&A, 443, 911

\bibitem[Chabrier et al.~1996]{chab96}
Chabrier G., Segretain L. and MŽra D., 1996, ApJ, 468, L21-L24

\bibitem[Chiba and Beers~2000]{chib00}
Chiba, M., and Beers, T.C., 2000, AJ, 119, 2843

\bibitem[Cr\'{e}z\'{e} et al.~2004]{crez04}
Cr\'{e}z\'{e}, M., Mohan, V., Robin, A.~C., Reyl\'{e}, C., McCraken, H.~J., 
Cuillandre, J.--C., Le~F\`{e}vre, O., Mellier, Y., 2004, A\&A, 426, 65

\bibitem[Cutri et al.~2003]{cutr03}
Cutri R. M., Skrustskie M. F., Van Dyk S. et al., 2003

\bibitem[Dahn et al.~2004]{dahn04}
Dahn, C.~C., Bergeron, P., Liebert, J., Harris, H.~C., Canzian, B., Leggett, 
S.~K., Boudreault, S. 2004, ApJ, 605, 400

\bibitem[Davies, King \& Ritter~2002]{davi02}
Davies, M.~B., King, A.~R., Ritter, H.~2002, MNRAS, 333, 469

\bibitem[Eichhorn \& Williams~1963]{eich63}
Eichhorn, H. and Williams, C.A.~1963, AJ, 68, 221

\bibitem[Eichhorn 1997]{eich97}
Eichhorn, H.1997, Astron. Astrophys., 327, 404

\bibitem[Flynn et al.~2003]{flyn03}
Flynn, C., Holopainen, J., Holmberg, J., 2003, MNRAS, 339, 817

\bibitem[Fuhrmann~2004]{fuhr04}
Fuhrmann K. 2004, Astron. Nact. 325:3-80

\bibitem[Gates et al.~2004]{gate04}
Gates, E.~et al.~2004, ApJ, 612, 129L

\bibitem[Hansen~1998]{hans98}
Hansen, B.~M.~S., 1998, Nature,  394, 860

\bibitem[Hansen~2003]{hans03}
Hansen, B.~M.~S.~2003, ApJ, 582, 915

\bibitem[Hansen and Liebert~2003]{hanl03}
Hansen, B.~M.~S. and Liebert, J. ~2003, ARA\&A, 41,465

\bibitem[Hambly et al.~2005]{hamb05}
Hambly, N. C., Digby, A. P., Oppenheimer, B. R., 2005, ASPC, 334, 113

\bibitem[Hawkins et al.~1998]{hawk98}
Hawkins, M.~R.~S., Ducourant, C., Jones, H.~R.~A. and Rapaport, M., 1998, MNRAS, 294, 505

\bibitem[Holopainen \& Flynn~2004]{holp04}
Holopainen, J., Flynn, C., 2004, MNRAS, 351, 721

\bibitem[Johnson and Soderblom~1987]{john87}
Johnson, D. R. H., Soderblom, D. R. 1987, AJ, 93, 864

\bibitem[Kilic et al.~2005]{kili05}
Kilic, M., Mendez, R.~A., von~Hippel, T., Winget, D.~E., 2005, ApJ, 633, 1126

\bibitem[Kowalski~2006]{kowa06}
Kowalski, P.~M.~2006, ApJ, 641, 488

\bibitem[Nale\.{z}yty et al.~2005]{made04}
Nale\.{z}yty, M., Madej, J., Althaus, L.~G.~2005, ASPC 334, 107

\bibitem[Mihalas and Binney~(1981)]{miha81}
Mihalas, D., Binney, J. 1981, "Galactic astronomy", second edition.

\bibitem[Monet et al.~1992]{mone92}
Monet D.G., Dahn C.C., Vrba F.J., Harris H.C., Pier J.R., Luginbuhl C.B., Ables H.D., 1992, AJ, 103, 638

\bibitem[Montiero et al.~2006]{mont06}
Montiero, H., Jao, W.--C., Henry, T., Subasavage, J., Beaulieu, T.~2006, ApJ,
638, 446

\bibitem[Oppenheimer et al.~2001]{oppe01}
Oppenheimer, B. R., Hambly, N. C., Digby, A. P., Hodgkin, S. T., Saumon, D. 2001, Science, 292, 698 (OHDHS)

\bibitem[Reid~2005]{reid05}
Reid, I.~N., 2005, ARA\&A, 43, 247

\bibitem[Robin et al.~1994]{robi94}
Robin, A., 1994, ApSS, 217, 163R

\bibitem[Salim et al.~2004]{sali04}
Salim, S., Rich, R.~M., Hansen, B.~M., Koopmans, L.~V.~E., Oppenheimer, B.~R., 
Blandford, R.~D., 2004, ApJ, 601, 1075

\bibitem[Torres et al.~2002]{torr02}
Torres, S., Garc\'{\i}a--Berro, E., Burket, A., Isern, J.~2002, MNRAS, 336, 971

\bibitem[Saumon \& Jacobson~1999]{saum99}
Saumon, D.; Jacobson, S. B.~1999,~ApJ, 511, L107

\bibitem[Silvestri et al.~2002]{silv02}
Silvestri, N.~M., Oswalt, T.~D., Hawley, S.~L.~2002, AJ, 124, 1118

\bibitem[Spagna et al.~2004]{spag04}
Spagna, A., Carollo, D., Lattanzi, M.~G., Bucciarelli, B.~2004, A\&A, 428, 451

\bibitem[Stetson et al.~1987]{stet87}
Stetson Peter B., 1987, PASP, 99, 191

\bibitem[Van Altena et al.~1995]{vana95}
Van~Altena, W.~F., Lee J.~T., Hoffleit E.~D.~1995, General Catalogue of 
Trigonometric Stellar Parallaxes, Fourth Edition, Yale University Observatory

\end{thebibliography}
\end{document}